\documentclass[12pt,amsmath,amssymb,]{revtex4}
\usepackage{graphics}
\usepackage{amsfonts}
\usepackage{amsmath}
\usepackage{epsfig}
\usepackage{epsf}
\usepackage{graphicx}
\usepackage{dcolumn}
\usepackage{bm}
\usepackage{color}
\usepackage{array}
\usepackage{mathrsfs}
\usepackage{amsmath}
\usepackage{braket}
\usepackage{indentfirst}
\usepackage{extarrows}
\usepackage{harpoon}

\newcommand {\sla}[1]{ #1 \!\!\!/}

\begin{document}

\title{ The $\sigma$ exchange effect in the Lamb shift of muonic hydrogen by two photons and two pions exchange}

\author{
Hai-Qing Zhou$^{1,2}$\protect\footnotemark[1]\protect\footnotetext[1]{E-mail: zhouhq@seu.edu.cn}\\
$^1$Department of Physics,
Southeast University, NanJing 211189, China\\
$^2$State Key Laboratory of Theoretical Physics, Institute of Theoretical Physics,\\
Chinese Academy of Sciences, Beijing\ 100190,\ P. R. China }
\date{\today}

\begin{abstract}
Based on the simple phenomenological $\sigma \pi \pi$, $\sigma N N$ and $\pi \pi \gamma$ interactions, we estimate the $\sigma$ exchange effect in the Lamb shift of muonic hydrogen. We at first calculate the effective couplings of $\sigma\mu\mu$ by two photons  and two pions exchange, then calculate the corresponding corrections to the energy shift of the 2S/2P states of muonic hydrogen. We find the correction to the energy shift of 2S state is about -14$\mu$eV by the current used parameters, which is about $44\%$ of the usual two photons exchange contribution, and is larger than the current experimental precision, and should be considered in the experimental analysis.
\end{abstract}

\maketitle

%%%%%%%%%%%%%%%%%%%%%%%%%%%%%%%%%%%%%%%%%%%%%%%%%%%%%%%%
\section{Introduction}
The puzzle of the proton size has attracted many interesting after the first precise measurement of the Lamb shift in muonic hydrogen \cite{Pohl2010}. Up to now, the CODATA2014 \cite{codata2014} gave $r_E=0.8751(61)$ fm based on the $ep$ scattering data and hydrogen data, while the recent precise measurement of the Lamb shift in muonic hydrogen confirmed the charge radius of proton $r_E=0.84087(39)$ fm\cite{Antognini2013-Science}. In the past a few years, many theoretical calculations on the energy spectrum of muonic hydrogen \cite{new-discussion-of-energy-spectrum,two-pion-exchange-to-energy-spectrum} and analysis on the $ep$ scattering data \cite{new-analysis-of-ep-scattering} have been done in the literatures, and these discussions showed the puzzle still persist and "how big is the proton" is still a serious question for us. And it is necessary for the theorists to give more careful discussions on the possible higher order's contributions in the Lamb shift.

\begin{figure}[htbp]
\center{\epsfxsize 2.0 truein\epsfbox{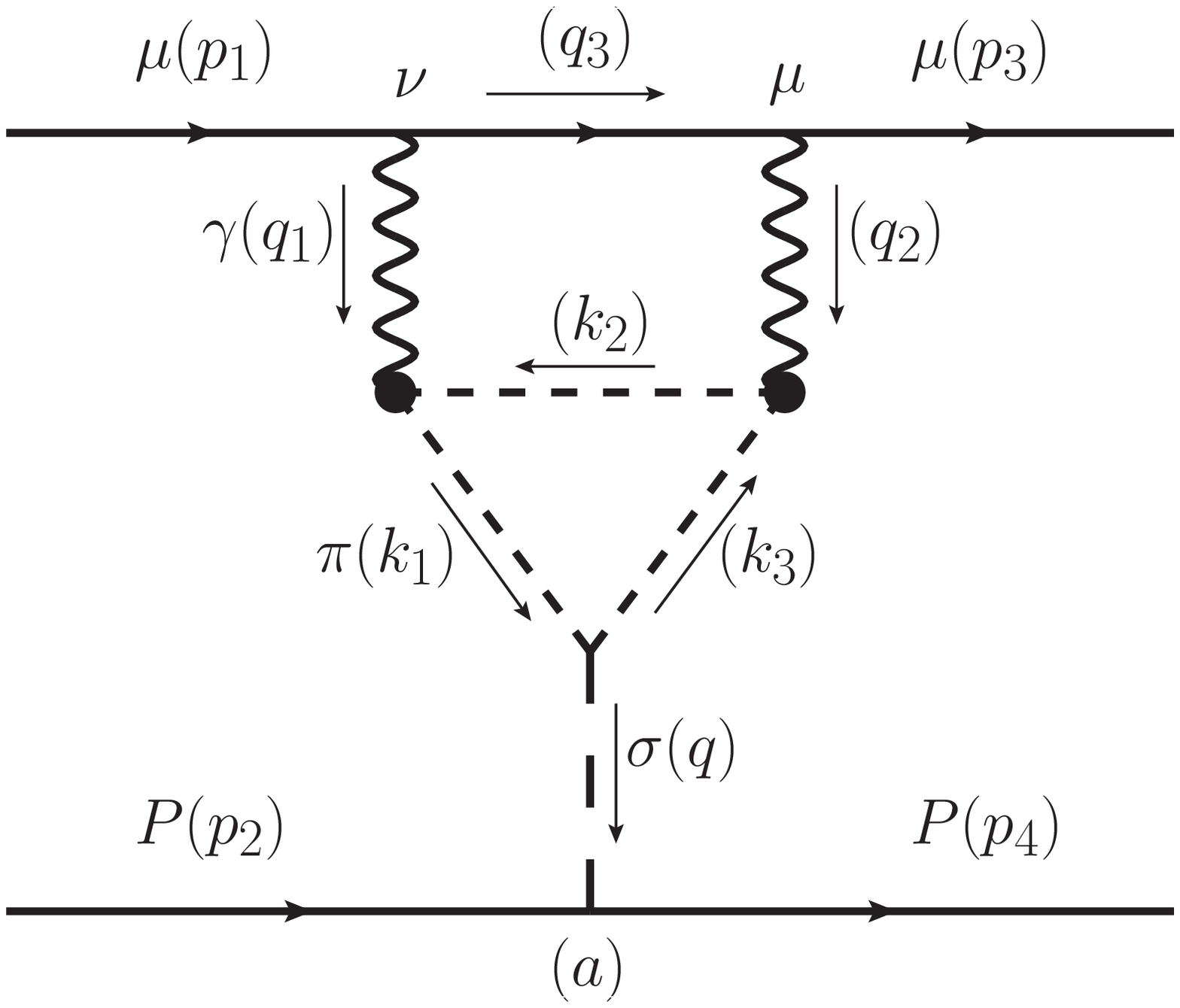}\epsfxsize 2.0 truein\epsfbox{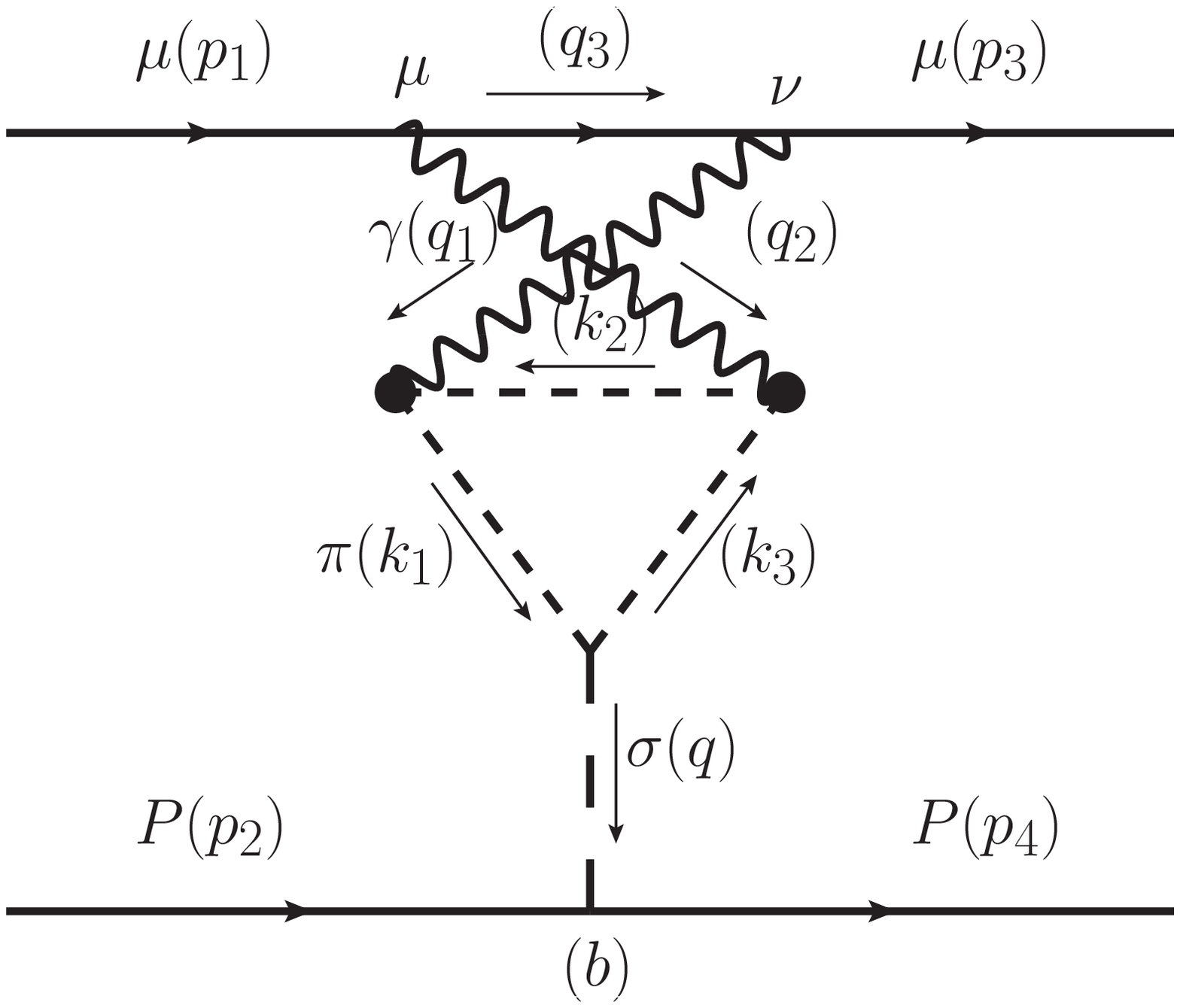}\epsfxsize 2.0 truein\epsfbox{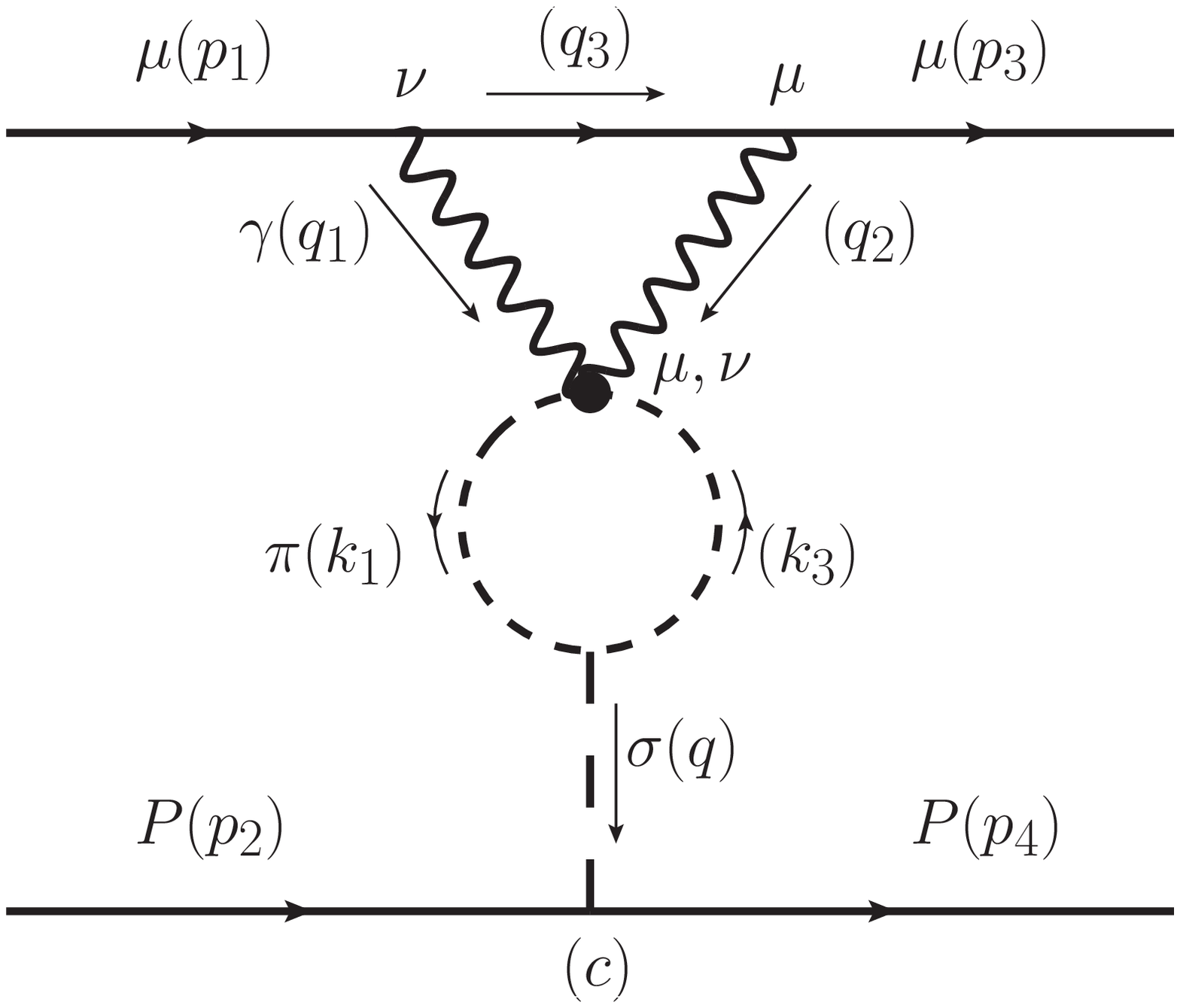}}
\caption{ The $\sigma$ exchange between the muon and the proton by two photons exchange, (a) the box like diagram; (b) the crossed-box like diagram; (c) the contact like diagram.}
\label{figure:sigma-exchange-diagram}
\end{figure}

In this work, we consider the  correction due to the scalar meson $\sigma$ exchange between the muon and the proton by two photons and two pions exchange which is showed as Fig. \ref{figure:sigma-exchange-diagram}. The properties of the scalar meson $\sigma$ has been discussed for a long time due to its large decay width and complex properties, and the recent PDG \cite{PDG2014} lists it with its pole mass as $\sqrt{s^{\sigma} _{pole}}=400-550 -i(200-350)$MeV. In the recent literatures \cite{one-pion-energy-spectrum-1,one-pion-energy-spectrum-2}, the one pion exchange corrections to the energy spectrum of muonic hydrogen are discussed and the contributions are found to be very small. At first glance, naively the pole mass of $\sigma$ is about 500 Mev which is much larger than the mass of $\pi$ and may give smaller contribution to the effective potential in the non-relativistic limit, and it is difficult to imagine that it can play its role in atomic physics. While we should note that actually the contributions from the electromagnetic form factors of proton (proton size) have been observed in atomic physics, and in some degree this effect is similar with the effect from a $\rho$ meson exchange between the muon and proton, due to vector meson dominance mechanism \cite{VMD-rho}. In the $NN$, $\pi N$ systems \cite{NN-piN-sigma}, it has been indicated that the $\sigma$ has large non-perturbative effective coupling with the $NN$ and $\pi \pi$. Different with the interaction between $\pi^0$ and two photons, there is no chiral anomaly for the electromagnetic interaction of $\sigma$.  Since the mass of $\sigma$ is smaller than $\rho$, we can expect the exchange of $\sigma$ may give certain contribution to the energy shifts of muonic hydrogen even its coupling to muon is by two photons. In this work, based on the simple effective interactions between $\sigma NN$, $\sigma \pi\pi$ and $\pi \pi\gamma$, we give an estimation on the $\sigma$ exchange effect  in the Lamb shift of the muonic hydrogen. The similar corrections in the Lamb shift of the hydrogen are very small due to the larger Bohr radius of hydrogen and are not discussed.

\section{Basic Formula}
Our start point is different with the calculations \cite{two-pion-exchange-to-energy-spectrum} based on the chiral effective theory .  In the chiral effective theory, the interactions are written down based on the chiral symmetry which is a property of QCD in the limit $m_{u,d}\rightarrow 0$, while since we are going to discuss the effects of strong interaction in the atomic physics level, the approximation $m_{u,d} \thickapprox 0$ is not a good approximation. And physically, the exist of $\pi,\sigma$ mesons and their interactions with proton and photons suggests we can construct the interaction including the scalar meson $\sigma$ directly in a phenomenological way.
We take the following effective interactions for $\sigma NN$, $\sigma \pi\pi$, and  $\pi\pi\gamma$ as our start point for our discussion.
\begin{eqnarray}
L_{\sigma NN} &=&  -g_{\sigma NN} \overline{\psi}_N\psi_N\sigma , \nonumber \\
L_{\sigma \pi\pi} &=& -\frac{1}{2}  g_{\sigma\pi\pi} \boldsymbol{\pi} \cdot \boldsymbol{\pi} \sigma , \nonumber \\
L_{\gamma\pi\pi} &=&   D_\mu^+ \pi^+ D^\mu \pi^-   ,
\label{effective-interactions}
\end{eqnarray}
where $\boldsymbol{\pi}=(\pi_1,\pi_2,\pi_3)$,$\pi^{\pm}=\frac{\sqrt{2}}{2}(\pi_1\pm i\pi_2)$,$\pi^0=\pi_3$, $D_\mu=\partial_\mu+ieA_\mu$ with $e=-|e|$.

The corresponding effective vertexes are expressed as
\begin{eqnarray}
&\Gamma_{\sigma NN} =-ig_{\sigma NN}, \Gamma_{\sigma\pi\pi} =-ig_{\sigma\pi\pi},  \nonumber \\
&\Gamma_{\gamma\pi\pi}^{\mu}  =-ie(p_1^\mu+p_2^\mu),\Gamma_{\gamma\gamma\pi\pi}^{\mu\nu}  = 2ie^2g^{\mu\nu} ,
\end{eqnarray}
with $\pm p_1,\pm p_2$ the momentums of the incoming  and outgoing $\pi^{\mp}$.
To include the effects from the electromagnetic structure of $\pi^{\pm}$, the electromagnetic form factor of pion is considered to describe the behavior of the effective coupling and we rewrite the effective vertexes as
\begin{eqnarray}
\widetilde{\Gamma}_{\gamma\pi\pi}^{\mu} &=&-ie(p_1^\mu+p_2^\mu)F(q^2),\nonumber \\
\widetilde{\Gamma}_{\gamma\gamma\pi\pi}^{\mu\nu} &=&ie^2g^{\mu\nu}F(q_1^2)F(q_2^2)
\end{eqnarray}
with $F(q^2)=-\Lambda^2/(q^2-\Lambda^2)$. We want to point out that the full amplitude is still gauge invariant by these effective couplings, and the including of the form factor also provides a natural regularization of the UV divergence. The choice of the parameter $\Lambda=0.77$GeV in this approach is not only consistent with the experimental data for the electromagnetic form factor of pion \cite{Pion-FormFactor}, but also give consistent result for $\pi^0\rightarrow e^+e^-$ \cite{PDG2014}. In the case we discussed, the suitable choice of the cut-off $\Lambda$ is also equivalent to the usual renormalization method by introducing a direct coupling interaction as counter term \cite{one-pion-energy-spectrum-2}. We will show this in the following .

By these interactions, the amplitude $\mu\rightarrow \mu \sigma^{*}$ by two photons and two pions exchange can be written down from the diagrams in the Fig.  \ref{figure:sigma-exchange-diagram},
\begin{eqnarray}
i\mathcal{M}_{\mu\rightarrow \mu\sigma^{*}} &=& i\mathcal{M}_{l\rightarrow l\sigma^{*}}^{(a)}+i\mathcal{M}_{l\rightarrow l\sigma^{*}}^{(b)}+i\mathcal{M}_{l\rightarrow l\sigma^{*}}^{(c)}\nonumber \\
&\equiv& -i g_{\sigma \mu\mu}(Q^2) \overline{u}(p_3,m_\mu)u(p_1,m_\mu),
\end{eqnarray}
with $m_\mu$ the mass of muon  and the $\mathcal{M}_{\mu\rightarrow \mu\sigma^{*}}^{(a,b,c)}$ refer to the amplitudes from the corresponding diagrams showed in Fig. \ref{figure:sigma-exchange-diagram}, $p_1,p_3$ are the momentums of the incoming and outgoing muon, $Q^2\equiv-q^2\equiv-(p_1-p_3)^2$, and the effective coupling $g_{\sigma \mu\mu}$ is also defined. In the Feynman gauge, the explicit expressions for the three amplitudes are written as
\begin{eqnarray}
i\mathcal{M}_{\mu\rightarrow \mu\sigma^{*}}^{(a)} &=&  \int \frac{d^4k_1 d^4k_2}{(2\pi)^8}\overline{u}(p_3,m_\mu)(-ie\gamma_{\mu}) \frac{i(\sla{q}_3+m_\mu)}{q_3^2-m_\mu^2+i\epsilon}(-ie\gamma_{\nu})u(p_1,m_\mu)\frac{-i }{q_1^2+i\epsilon}
\frac{-i }{q_2^2+i\epsilon} \nonumber \\
&&\frac{i}{k_1^2-m_\pi^2+i\epsilon}\frac{i}{k_2^2-m_\pi^2+i\epsilon}\frac{i}{k_3^2-m_\pi^2+i\epsilon} \widetilde{\Gamma}_{\gamma\pi\pi }^{\mu}(q_1) \widetilde{\Gamma}_{\gamma\pi\pi}^{\nu}(q_2) \Gamma_{\sigma \pi\pi}, \nonumber \\
i\mathcal{M}_{\mu\rightarrow \mu\sigma^{*}}^{(b)} &=&  \int \frac{d^4k_1 d^4k_2}{(2\pi)^8}\overline{u}(p_3,m_\mu)(-ie\gamma_{\nu}) \frac{i(\sla{q}_3+m_\mu)}{q_3^2-m_\mu^2+i\epsilon}(-ie\gamma_{\mu})u(p_1,m_\mu)\frac{-i }{q_1^2+i\epsilon}
\frac{-i }{q_2^2+i\epsilon} \nonumber \\
&&\frac{i}{k_1^2-m_\pi^2+i\epsilon}\frac{i}{k_2^2-m_\pi^2+i\epsilon}\frac{i}{k_3^2-m_\pi^2+i\epsilon} \widetilde{\Gamma}_{\gamma\pi\pi }^{\mu}(q_1) \widetilde{\Gamma}_{\gamma\pi\pi}^{\nu}(q_2) \Gamma_{\sigma \pi\pi}, \nonumber \\
i\mathcal{M}_{\mu\rightarrow \mu\sigma^{*}}^{(c)} &=&  \int \frac{d^4k_1 d^4k_2}{(2\pi)^8}\overline{u}(p_3,m_\mu)(-ie\gamma_{\mu}) \frac{i(\sla{q}_3+m_\mu)}{q_3^2-m_\mu^2+i\epsilon}(-ie\gamma_{\nu})u(p_1,m_\mu)\frac{-i }{q_1^2+i\epsilon}
\frac{-i }{q_2^2+i\epsilon} \nonumber \\
&&\frac{i}{k_1^2-m_\pi^2+i\epsilon}\frac{i}{k_2^2-m_\pi^2+i\epsilon}\widetilde{\Gamma}_{\gamma\pi\pi }^{\mu\nu} \Gamma_{\sigma \pi\pi},
\label{eq:effective-couplings}
\end{eqnarray}
where $m_\pi$ is the mass of pion, $q_{1,2,3}$ and $k_{1,2}$ are the corresponding momentums of photons and pions showed in the corresponding diagrams of Fig. \ref{figure:sigma-exchange-diagram}.  We use the package Feyncalc9 \cite{FeynCalc9} to simplify the Dirac matrix and use the FIESTA4.1 \cite{FIESTA} to do the numerical calculation for the two loop integration. Since we are interesting in the corrections to the energy shifts of muonic hydrogen, the behavior of the couplings $g_{\sigma\mu\mu}(Q^2)$ in the non-relativistic approximation is important to such corrections, we expand the effective couplings at $Q^2=0$ and
we find it can be expressed in a normal form as
\begin{eqnarray}
g_{\sigma \mu\mu} &=& g_{\sigma\pi\pi}[c_1+c_2Q ].
\label{eq:2}
\end{eqnarray}
In the practical calculation, we find that only $c_1$ is dependent on the parameter $\Lambda$, this can be understood naturally, since the parameter $\Lambda$ plays the role
as the  regularization parameter which is corresponding to the $Q^2$ independent counter term in the effective interactions, this means in our case, the result by introducing the form factor is equivalent to introduce a direct coupling of meson-muon-muon as counter term.

By the effective coupling $g_{\sigma \mu\mu}$, we can use the quasipotential method by matching the amplitude from the effective interactions in quantum field theory and that from the effective non-relativistic potential in quantum mechanism as similar as the case of pion exchange between nucleon nucleon.
\begin{eqnarray}
i\mathcal{M}_{\mu p\rightarrow \mu p} &=& \overline{u}(p_3,m_\mu)(-ig_{\sigma\mu\mu})u(p_1,m_\mu) u(p_4,m_N)(-ig_{\sigma NN})u(p_2,m_N)\frac{i}{q^2-m_\sigma^2+i\epsilon}
\nonumber \\
&\stackrel{NR}{\approx}&   -i  \langle f|V_{lp}|i\rangle.
\end{eqnarray}
%$\xlongrightarrow[approximation]{NR } $
where $m_N,m_\sigma$ are the masses of proton and $\sigma$ respectively, and $NR$ refers to the non-relativistic approximation. The effective potential in the momentum space and coordinate space can be expressed as
\begin{eqnarray}
V_{\mu p}^{\sigma}(|\overrightharp{q}|) & = & -\frac{g_{\sigma \mu\mu}g_{\sigma NN}}{m_\sigma^2+\overrightharp{q}^2} = -\frac{g_{\sigma \pi\pi}g_{\sigma NN}}{m_\sigma^2+\overrightharp{q}^2}(c_1+c_2q), \nonumber \\
V_{\mu p}^{\sigma}(r) & = & \int\frac{d^3\vec{q}}{(2\pi)^3}e^{i\overrightharp{q}\cdot\overrightharp{r}} V_{\mu p}^{\sigma}(|\overrightharp{q}|),
\label{Eq-effective-potential}
\end{eqnarray}
with $\overrightharp{q}^2=-q^2=Q^2$ and $r=|\vec{r}|$. The corrections to the energy level of $2S/2P$ states in the perturbative theory are
\begin{eqnarray}
\Delta E_{2S,2P} = \int _{0}^{\infty}\psi_{2S,2P}^2 V_{\mu p}^{\sigma}(r)r^2 dr
\end{eqnarray}
and can be calculated directly as
\begin{eqnarray}
\Delta E_{2S} &=&-g_{\sigma\pi\pi}g_{\sigma NN}\Big\{\frac{2x^2+1}{16\pi a_{u}(x+1)^4}c_1+\frac{m_\sigma^2}{8\pi^2x^2(1-x^2)^4}\nonumber\\
& &\times \big[-9x^6+7x^4+x^2+4(2x^6+3x^4+x^2)logx+1\big]c_2 \Big\} \nonumber \\
\Delta E_{2P} &=&-g_{\sigma\pi\pi}g_{\sigma NN}\Big\{\frac{1}{16\pi a_{u}(x+1)^4}c_1+\frac{m_\sigma^2}{24\pi^2x^2(1-x^2)^4}\nonumber\\
& &\times\big[-x^6+9x^4+9x^2+12x^2(x^2+1)logx+1\big]c_2\Big\}
\label{Eq:energy-correction}
\end{eqnarray}
with $x=a_{u}m_{\sigma}$, $a_{u}$ the Bohr radius of muonic hydrogen.
%And after the integration the numerical results for the energy shift are
%\begin{eqnarray}
%\Delta E_{2S} = \int  \nonumber \\
%\Delta E_{2P} = \int  \nonumber \\
%\label{Eq-effective-potential}
%\end{eqnarray}

%\Section{The Perturbative Potential From The Effective Coupling Of Pion Ee And Pion Nn }

\section{Parameters in the effective interaction}

To give a numerical estimation to the $\Delta E_{2S,2P}$, we should at first give an estimation to the coupling parameters $g_{\sigma\pi\pi}$, $g_{\sigma NN}$ and the mass of $\sigma$. For the mass of $\sigma$ in the propagator, we approximately take it as the real part of its pole mass to estimate the contribution, which means we directly take $m_{\sigma}=0.45$ GeV as an example. By the effective interactions Eq.(\ref{effective-interactions}), the decay width $\Gamma(\sigma\rightarrow \pi\pi)$ can be calculated directly which can be expressed as
\begin{eqnarray}
\Gamma(\sigma\rightarrow \pi\pi) &=&\frac{3g_{\sigma\pi\pi}^2\sqrt{m^2_\sigma-4m^2_\pi}}{32\pi m^2_\sigma}.
\end{eqnarray}
If we take the decay width of $\sigma$ as $\Gamma_{\sigma}=0.5$ GeV, we can get $|g_{\sigma \pi\pi}|=3.1$ GeV, which is close to the values presented in the literatures \cite{sigma-pi-pi}. For $g_{\sigma NN}$, we directly take it as $|g_{\sigma NN}|=14$ which is given in many literatures \cite{NN-piN-sigma}. The relative phase  between the $g_{\sigma NN}$ and $g_{\sigma \pi\pi}$ (0 or $\pi$) is a little difficult to be determined. In principle,, this phase can be estimated from the phase of the effective coupling of $\sigma\rightarrow\gamma\gamma$,  here we do not go to discuss this question in detail and just assume the relative phase is 0.

\section{Numerical results, discussion and conclusion }
\begin{figure}[htbp]
\center{\epsfxsize 5.0 truein\epsfbox{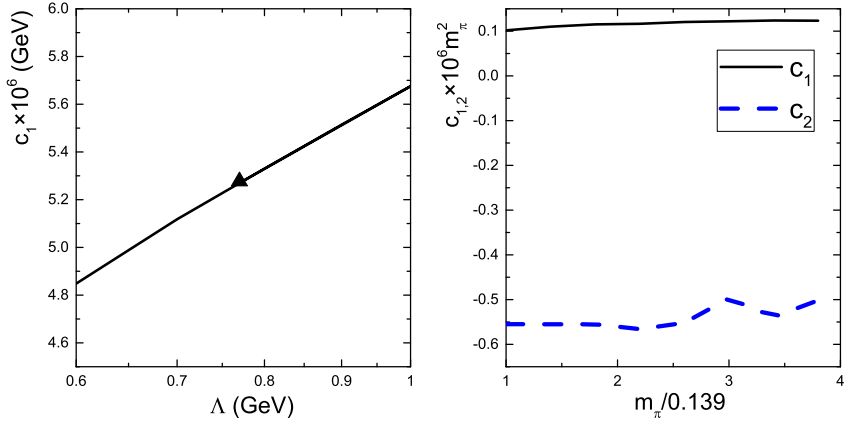}}
\caption{(a) The dependence of $c_{1,\mu}$ on the parameter $\Lambda$; (b) the dependence of $c_1$ on the mass of pion $m_\pi$.}
\label{figure:c1-dependence}
\end{figure}

Numerically, we get $c_2=-29\times10^{-6}$ GeV$^{-1}$ which is only dependent on the physical mass of pion and muon. And the  dependence of $c_1$ on the cut-off $\Lambda$ and $m_{\pi}$ are presented in the Fig. \ref{figure:c1-dependence}.
From the left panel of Fig. \ref{figure:c1-dependence}, we see that $c_{1}$ is not sensitive on the parameter $\Lambda$, which means the uncertainty from the electromagnetic structure of pion is small. From the right panel of Fig. \ref{figure:c1-dependence} , we see $c_{1,2}$ decrease quickly  when the masses of pion $m_{\pi}$ increase (the behavior is like $c_{1,2}\propto 1/m_\pi^2$), which means the contributions from other $0^{-+}$ measons with higher mass can be neglected.

By Eq. (\ref{Eq:energy-correction}) and the numerical results of $c_{1,2}$, the last numerical results for the energy shifts of muonic hydrogen can be gotten easily which are
\begin{eqnarray}
&\Delta E_{2S}= -14 \mu eV, \nonumber \\
&\Delta E_{2P}=-0.0029 \mu eV,
\end{eqnarray}
and almost all the contributions are from the $c_1$ term for $2S$ state and $c_2$ term for $2P$ state. From the Fig. 2, we can see the main uncertainty in our calculation is from the input values for the mass of $\sigma$,  the couplings $g_{\sigma NN}$  and $g_{\sigma\pi\pi}$ .  The precise determine of these values are difficult due to the nonperturbative properties of QCD. In our discussion, we just take the center values from the literatures. And naively we can expect that all these values have at least 10\%  uncertainty,  and the final result have at least 30\% uncertainty, while the order of the last contribution is certain.
In the literatures, the two photons exchange correction to the Lamb shift usually is taken as $33$ $\mu$eV \cite{Carlson2015}. The contribution  $-14$ $\mu$eV is about 44\% of such correction and about 4\% of the current discrepancy ($0.316$ meV) \cite{Jentschura2011}. The order of this correction is still larger than the current experimental precision and should be considered in the experimental analysis.

%We also present the dependence of $\Delta E_{2S}$ on the mass of $m_{\sigma}$ in the Fig.4 where we see the contributions is about from  to .

In summary, we discuss the $\sigma$ exchange effect in muonic hydrogen by two photons and two pions exchange and find it gives about $-14$ $\mu$eV to the Lamb shift of muonic hydrogen by the present used parameters and this correction should be considered under the current experimental precision.

\section{Acknowledgments}
This work is supported by the  National Natural Science Foundations of China
under Grant No. 11375044 and in part by the Fundamental Research Funds for the Central Universities under Grant No. 2242014R30012. The author thanks Wen-Long Sang and A.V. Smirnov for the help on FIESTA, and thank Bing-Song Zou, Zhi-Yong Zhou and Feng-Kun Guo for the helpful discussion.

\end{document}